# Energy Spectra vs. Element Abundances in Solar Energetic Particles and the Roles of Magnetic Reconnection and Shock Acceleration

**Donald V. Reames** (https://orcid.org/0000-0001-9048-822X)

Institute for Physical Science and Technology, University of Maryland, College Park, MD 20742-2431 USA, email: dvreames@gmail.com

**Abstract** We reexamine the relationship between energy spectral indices and element abundance enhancements in solar energetic particle (SEP) events at energies of a few MeV amu$^{-1}$. We find a correlated behavior only in the largest gradual SEP4 events when all ions are accelerated from the ambient coronal plasma by shock waves driven by fast, wide coronal mass ejections (CMEs). This correlated abundance behavior can track complex time variations in the spectral indices during an event. In other (SEP3) events, CME-driven shock waves, days apart, sample seed particles from a single pool of suprathermal impulsive ions contributed earlier. Of the smaller, Fe-rich, impulsive SEP events, previously related to magnetic reconnection in solar jets, over half are subsequently reaccelerated by CME-driven shock waves (SEP2) causing typical ion intensities to have a 64% correlation with shock speed. In these SEP2 events, onset of shock acceleration is signaled by a new component in the abundances, large proton excesses. The remaining SEP1 events lack evidence of shock acceleration. However, for all these events (SEP1 – SEP3) with abundances determined by magnetic reconnection, energy spectra and abundances are decoupled.

## 1. Introduction

After sixty years of study it has become clear that acceleration of solar energetic particles (SEPs) is quite complex (e.g. Reames, 2021b, 2021c). Two major processes (Reames, 1988, 1995b, 1999, 2013, 2015; Gosling 1993), magnetic reconnection and shock acceleration, determine the primary physics, with the assistance of wave-particle interactions in some cases (e.g. Temerin and Roth, 1992; Ng, Reames, and Tylka, 1999, 2003). Acceleration of small "impulsive" SEP events appears to occur at sites of magnetic reconnection (Drake et al., 2009) on open magnetic field lines in solar jets (Bučík et al., 2018a, 2018b), or on closed field lines in flares (Mandzhavidze, Ramaty, and Kozlovsky, 1999; Murphy et al., 1991, 2016), producing distinctive particle enhancements of $^3$He (e.g. Serlemitsos and Balasubrahmanyan, 1975; Temerin and Roth, 1992; Mason, 2007; Bučík, 2020; Reames 2021d) and of heavy elements (Mason et al., 1986; Reames, Meyer, and von Rosenvinge, 1994) up to Au and Pb (Reames, 2000; Reames and Ng, 2004; Mason et al., 2004; Reames, Cliver, and Kahler, 2014a, 2014b, 2015). Larger "gradual" SEP events are accelerated by shock waves driven out from the Sun by wide, fast coronal mass ejections (CMEs; Kahler, et al., 1984; Mason, Gloeckler, and Hovestadt, 1984; Reames, Barbier, and Ng, 1996; Zank, Rice, and Wu, 2000; Cliver, Kahler, and Reames, 2004; Lee, 2005; Lee, Mewaldt, and Giacalone, 2012; Desai and Giacalone, 2016). Complexity is introduced when small but fast CMEs from jets (Kahler, Reames, and Sheeley, 2001; Nitta, et al., 2006; Reames, Cliver, and Kahler, 2014) produce shock waves that reaccelerate earlier SEP ions, or when large shock waves in gradual events encounter pools of collected suprathermal ions from many small jets (Desai et al., 2003; Bučík et al., 2014, 2015; Chen et al., 2015; Bučík, 2020).

Despite this complexity, we commonly find simple power-law behavior in intermediate-energy spectra and in the dependence of element abundances on the mass-to-charge ratio, $A/Q$, of the ions in gradual (Breneman and Stone 1985; Reames 2016, 2021b) and impulsive (Reames, Cliver, and Kahler, 2014a, 2014b) SEP events. In the largest shock-driven SEP events these dependences even seem to be coupled (Reames, 2020b, 2021a). Reames (2020a) has identified four paths to produce the observed SEP element abundance patterns:

(i) SEP1: "pure" impulsive events from magnetic reconnection in solar jets with enhancements of $^3$He and heavy elements, but no shock acceleration,

(ii) SEP2: impulsive events with reacceleration by a local CME-driven shock,

(iii) SEP3: gradual events dominated by pre-accelerated impulsive ions at $Z > 2$,

(iv) SEP4: gradual events dominated by seed ions from the ambient coronal plasma.

An aid to the discrimination of these categories emerged when power-law fits to element abundance enhancements vs. $A/Q$ were extended from elements with $Z > 2$ down to protons at $A/Q = 1$. Apparently, a large excess of protons means that the seed particles for shock acceleration included protons from the ambient coronal plasma plus the enhanced suprathermal impulsive ions of higher $Z$, in both impulsive (Reames 2019a) and gradual (Reames 2019b) SEP events. Thus, enhanced heavy ions plus excess protons was a signature of shock acceleration of complex seeds in SEP2 and SEP3 events.

At the highest energies, protons above $\approx 1$ GeV can form a nuclear cascade through the Earth's atmosphere so these SEP events are detectable at ground level when their intensities become comparable with those of the ubiquitous galactic cosmic rays (GCRs). These ground-level enhancement events (GLEs) were the earliest indication of SEPs (Forbush, 1946), but they provide no information on SEP abundances. Worse, the energy spectra of SEP events were found to break downward above about 10 MeV (Tylka et al., 2005; Tylka and Dietrich, 2009; Mewaldt et al., 2012; see also Gopalswamy et al., 2012), and since that break introduces a new dependence upon rigidity or $A/Q$ (Li et al., 2009; Zhao et al., 2016), it brings new physics into the element abundances at high energies. Below $\approx 1$ MeV amu$^{-1}$, particle transport in large SEP events is often impeded by waves generated by the outward-streaming protons themselves, which differentially scatter the ions and flatten the spectra (Reames and Ng, 2010; Ng, Reames, and Tylka, 1999, 2003, 2012). Thus we are left with the region of about $1 - 10$ MeV amu$^{-1}$ where the energy spectra and abundances retain the most information on their relatively unmodified acceleration spectra at their source.

The variable we often call "energy", $E$, in MeV amu$^{-1}$, is actually a function of only ion velocity $E = \mathcal{E}/A = M_u(\gamma - 1) \approx \frac{1}{2} M_u \beta^2$, where $\mathcal{E}$ is the total kinetic energy, $A$ is the atomic mass, the mass unit $M_u = m_u c^2 = 931.494$ MeV, $\gamma = (1 - \beta^2)^{-1/2}$, and $\beta = v/c$ is the particle velocity relative to that of light, *c*. The magnetic rigidity, or momentum per unit charge, is $P = pc/Qe = M_u \beta\gamma A/Q$ in units of MV, where *p* is the momentum. Since power laws so commonly represent spectra and abundances, we can represent the observed intensity spectra as

$$j(A/Q, E) = k\, (A/Q)^a\, E^b, \tag{1}$$

where *k* is a constant independent of *A/Q* and *E*. In general the parameters *a* and *b* are unrelated, especially when abundance enhancements are predetermined by one process and the spectra by a subsequent one. However, an underlying result has been found (Reames 2020b, 2021a) for the case of pure shock acceleration in SEP4 events, where,

$$b = a/2 - 2. \tag{2}$$

This equation relates the energy spectrum of one ion to the selection and acceleration of other species of ions from the same source. When Equation 2 is valid, measurement of the relative abundances of heavy ions can determine the proton spectral index, and conversely. Equation 2 can be disrupted by wave excitation during transport, which scatters and delays low-rigidity He, for example, much more than higher-rigidity Fe of the same velocity or *E*.

The purpose of this article is to consider a different perspective on the time variation of the correlated enhancement and suppression of energy spectra and abundances in SEP4 events suggested by Equation 2, and the decoupled variation in the SEP3 events. We also present evidence for the persistence of the pools of seed populations for many days while they contribute to multiple large SEP3 events. Especially, we test the meaning and the power of the proton excess, in the element abundance patterns, to distinguish both SEP2 and SEP3 events, whereas shock speed and source plasma temperature have been considered previously (Reames 2021a), and we compare different measures of shock acceleration in impulsive SEP events. How extensive is shock acceleration?

We use SEP measurements from the Wind spacecraft near Earth. Wind provides relative abundances of the elements H, He, C, N, O, Ne, Mg, Si, S, Ar, Ca, and Fe, and also groups of heavier elements up to Pb from the Low-Energy Matrix Telescope (LEMT; von Rosenvinge et al., 1995; Reames, 2021b). Abundances are primarily from the 3.2–5 MeV amu$^{-1}$ interval on LEMT, although H is only available near 2.5 MeV amu$^{-1}$. Abundance enhancements are measured relative to the average SEP abundances listed by Reames (2021b; see also Reames, 1995a, 2014, 2018a, 2020a). Event lists are available for impulsive (Reames, Cliver, and Kahler, 2014a) and gradual (Reames 2016) events. Wind/LEMT data are available at https://cdaweb.gsfc.nasa.gov/sp_phys/.

## 2. Power Laws in *A/Q* and Temperatures

While energy spectral indices are broadly understood and examples of the quality of fits to spectral indices in these SEP events have been shown (Reames, 2021a), fits to abundance enhancements vs. *A/Q*, when *Q* varies with the coronal temperature, may be less obvious. We determine *Q* by selecting the temperature which gives the best-fit power law of element abundance enhancements vs. *A/Q*. This process has been used to obtain power-law fits for impulsive (Reames, Cliver, and Kahler, 2014b, 2015; Reames, 2019a) and gradual (Reames, 2016, 2019b) SEP events and has been discussed in a review (Reames, 2018b) and even a textbook (Reames, 2021b).

Impulsive SEP events are all significantly "Fe-rich", having a positive power-law dependence on average as $(A/Q)^{3.64\pm0.15}$ (Reames, Cliver, and Kahler, 2014a) that seems to be a characteristic of acceleration in the magnetic reconnection regions (e.g. Drake et al., 2009). Gradual SEP events can also have rising power-laws in *A/Q*, either because they reaccelerate suprathermal impulsive ions from the seed population in SEP3 events or because of strong scattering by proton-amplified Alfvén waves (Stix, 1992; Ng and Reames, 1994; Ng, Reames, and Tylka, 1999, 2001, 2003; Reames and Ng, 2010) early in large SEP4 events, which preferentially retards lower-rigidity, low-*Z* ions more than higher-rigidity, high-*Z* ions, so the power of *A/Q* decreases with time. Typical large events illustrating each behavior are shown in Figure 1. Figure 1c illustrates the definition of proton excess in the last time interval, which has been taken as evidence of two components of the seed population, the accelerated ambient ion population running approximately along the dashed lines (suitably extended), while the impulsive suprathermal contribution follows the solid lines (see e.g. Figure 6 in Reames 2021c).

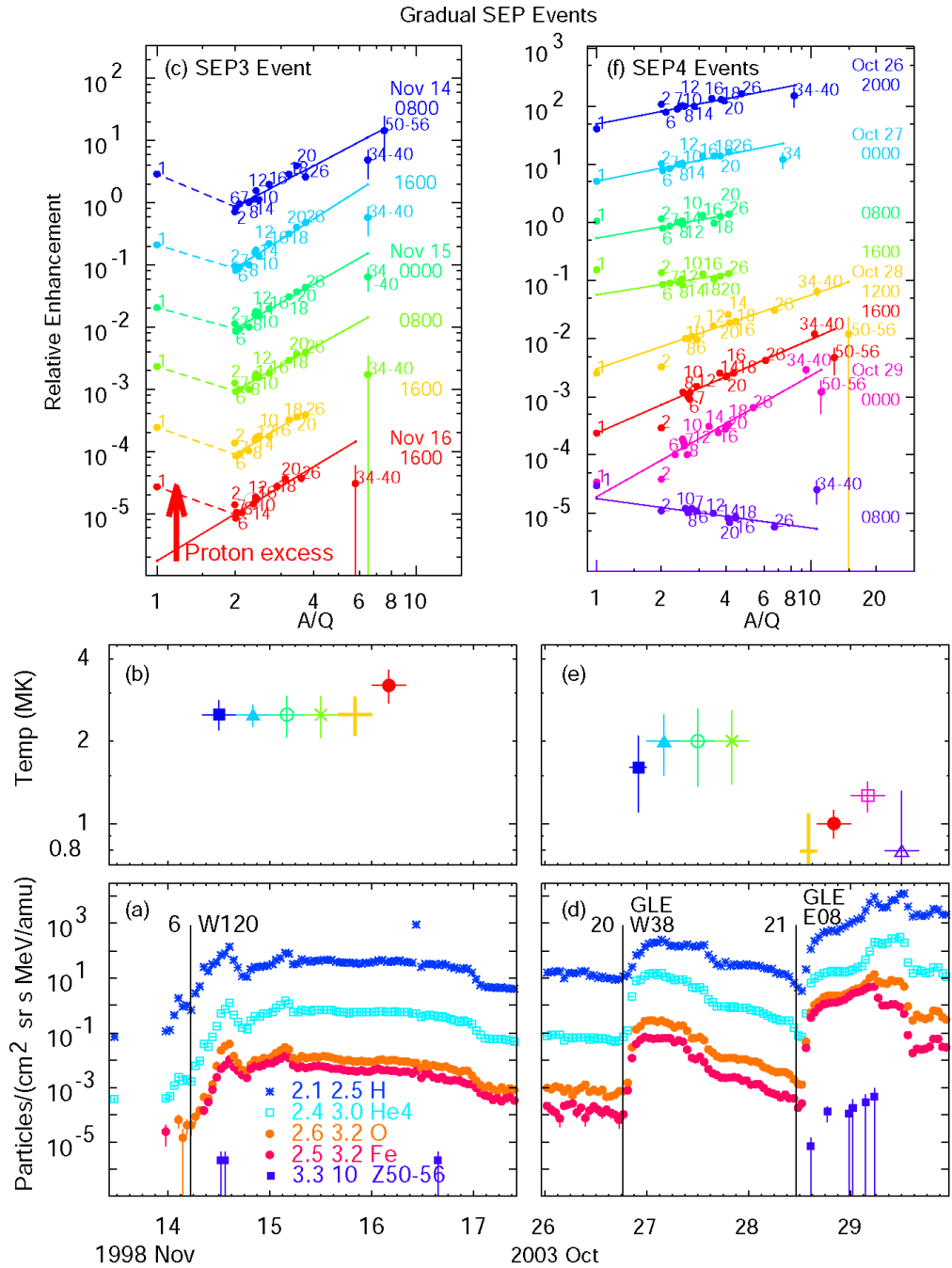

**Figure 1** The SEP3 event in November 1998 (left panels) is compared with two large SEP4 events of October 2003 (right panels). Panels (**a**) and (**d**) Compare element time profiles, (**b**) and (**e**) compare best-fit plasma temperatures, and (**c**) and (**f**) compare best-fit power-law abundances vs. $A/Q$ in each time-interval with onset listed and colors corresponding to temperature intervals below. Vertical lines in panels a and d mark onset times with event number from Reames (2016), and source location. Proton excess is defined in panel c where the dashed lines suggest the maximum contributions of seed particles from ambient plasma.

SEP3 events can also be distinguished by their best-fit source plasma temperature, shown in this case as 2.5 MK, typical of the impulsive ions (Reames, Cliver, and Kahler, 2014b) and recently confirmed by the differential emission measure in jets (Bučík et al., 2021) that produced these seed components. Alternatively, SEP3 events can be identified by the proton excess that appears to identify a dual power-law seed population. For the SEP4 events in Figure 1f, all the elements tend to support single power laws from a single dominant seed population.

In the next section, we discuss SEP4 events, followed with SEP3 events, and then work our way down to distinguishing SEP1 and SEP2 events.

## 3. SEP4 Events

Since our purpose is to compare energy spectra with abundances, we show a more generic SEP4 event in Figure 2. In Figure 2c the derived temperatures are lower, more typical of ambient coronal plasma, and in Figure 2d, the abundances decline with *A/Q*, showing a lack of either extreme turbulence or impulsive suprathermals. Nevertheless, this event is still a GLE with hard high-energy spectra. Figure 2b shows that the powers of *A/Q* decline with time, and, more importantly, the expected power of *E*, derived from Equation 2, falls among the observed energy spectral indices for He, O and Fe.

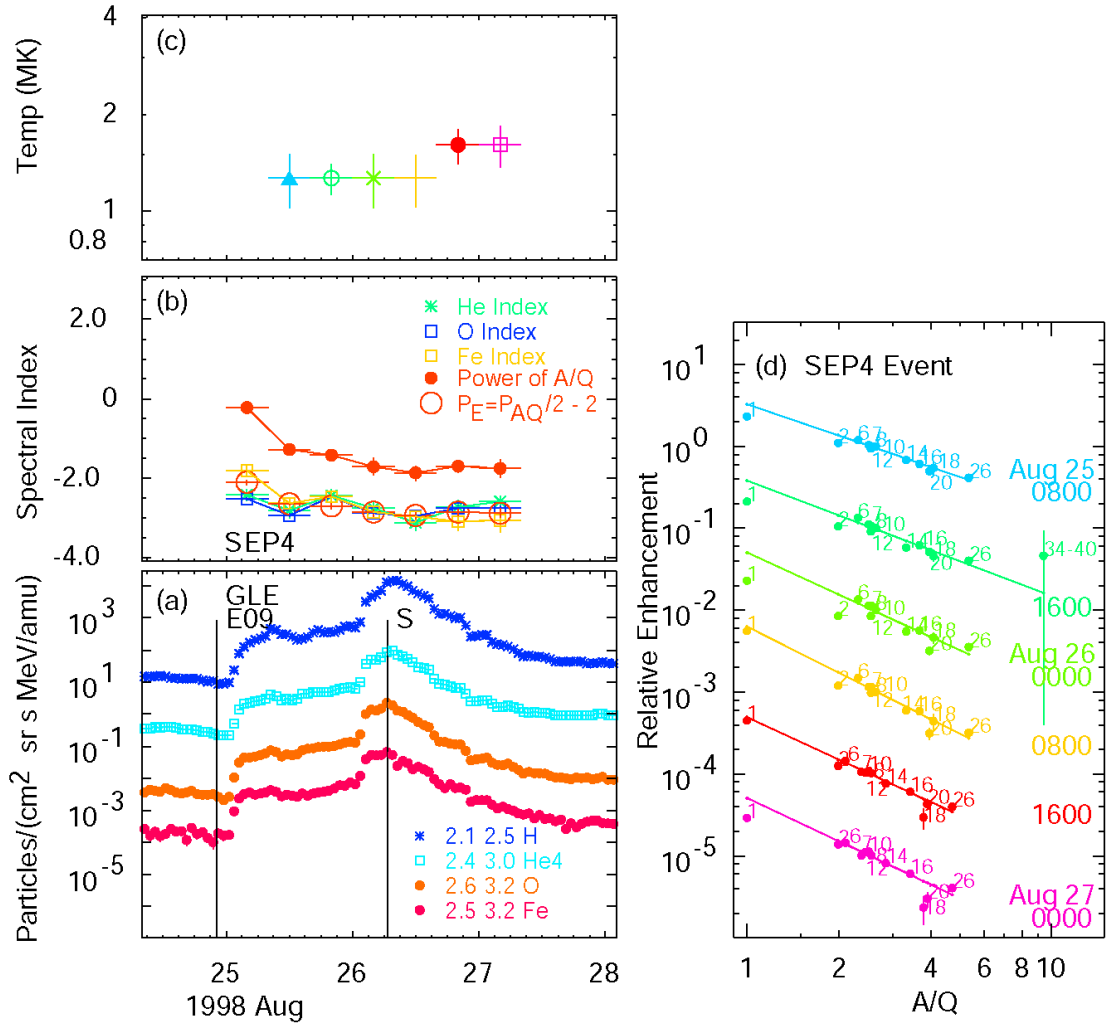

**Figure 2** For the SEP4 event of 24 August 1998 we show (**a**) element time profiles, (**b**) observed spectral indices for He, O, and Fe compared with the power of *A/Q* (solid red circle) and the expected spectral index derived from Equation 2 (open red circles), (**c**) derived plasma temperatures, and (**d**) best-fit power-law abundances vs. *A/Q* in each time-interval.

Figure 3 shows two very large SEP4 events, one a GLE. Figures 3b and 3e show that the spectra of He are flatter than those of Fe early in the events, known from the effects of the wave generation and the "streaming limit" (e.g. Reames and Ng, 2010). The power of *A/Q* also varies from strongly positive early to mildly negative after shock passage (as for the GLEs in Figure 1f). However, the expected spectral indices derived from Equation 2 follow the trend of the observed spectral indices in Figure 3b and 3e.

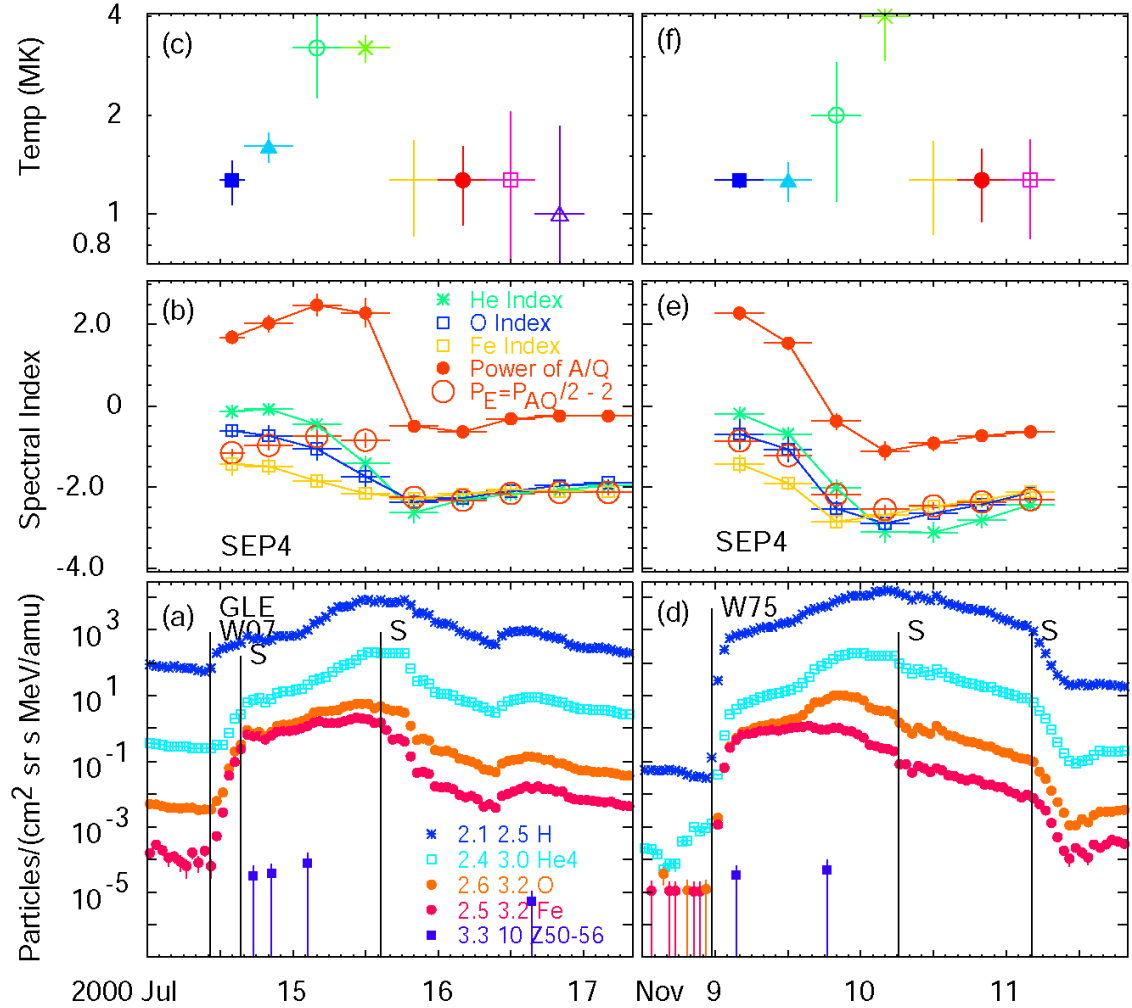

**Figure 3** Shows two large SEP4 events where (**a**) and (**d**) compare element time profiles, (**b**) and (**e**) show observed spectral indices for He, O, and Fe along with the power of *A/Q* (solid red circles) and the expected spectral indices derived from Equation 2 (open red circles), and (**c**) and (**f**) compare best-fit temperatures in each time-interval (for detailed power-law fits see Reames, 2020b). Temperatures are poorly defined when abundance powers approach zero, where any *A/Q* (from any *T*) gives the same enhancement, or at shock peaks where high intensities may begin to saturate LEMT.

As final examples of gradual SEP4 events at an opposite extreme we show the events in Figure 4. For the event on the left, the power of *A/Q* and the derived power of *E* both approach -4 in Figure 4b, as *a* = *b* = -4 is a solution to Equation 2. For the moderate event on the right, Figure 4e shows that the power of *A/Q* swings from positive early to negative, with temperatures in Figure 4f being undefined as it crosses through zero.

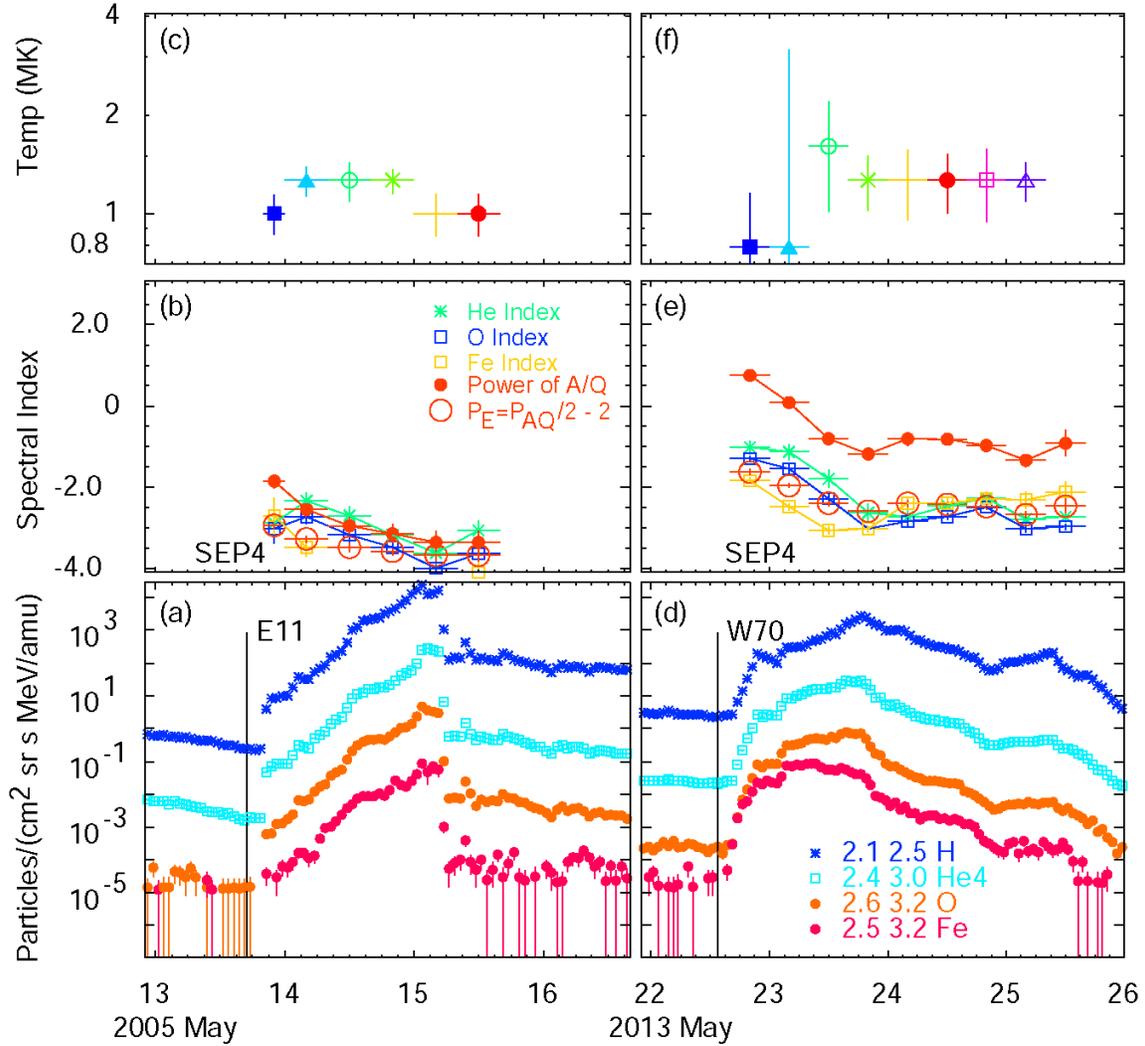

**Figure 4** Two smaller SEP4 events where (**a**) and (**d**) compare element time profiles, (**b**) and (**e**) show observed spectral indices for He, O, and Fe along with the power of *A/Q* (solid red circles) and the expected spectral index derived from Equation 2 (open red circles), and (**c**) and (**f**) compare best-fit temperatures in each time-interval.

## 4. SEP3 Events

Figure 5 shows the variation of parameters during a period containing two SEP3 events. Panel 5d is included as a reminder of the proton excess for these SEP3 events during the GLE event of 7 November 1997. For both of these events shown in Figure 5b, the powers of *E* derived via Equation 2 seem to have nothing to do with the observed spectral indices of He, O, or Fe, i.e. the abundances are independent of the spectra. Presumably, the abundances from pools of impulsive suprathermal ions attain their spectra in shocks.

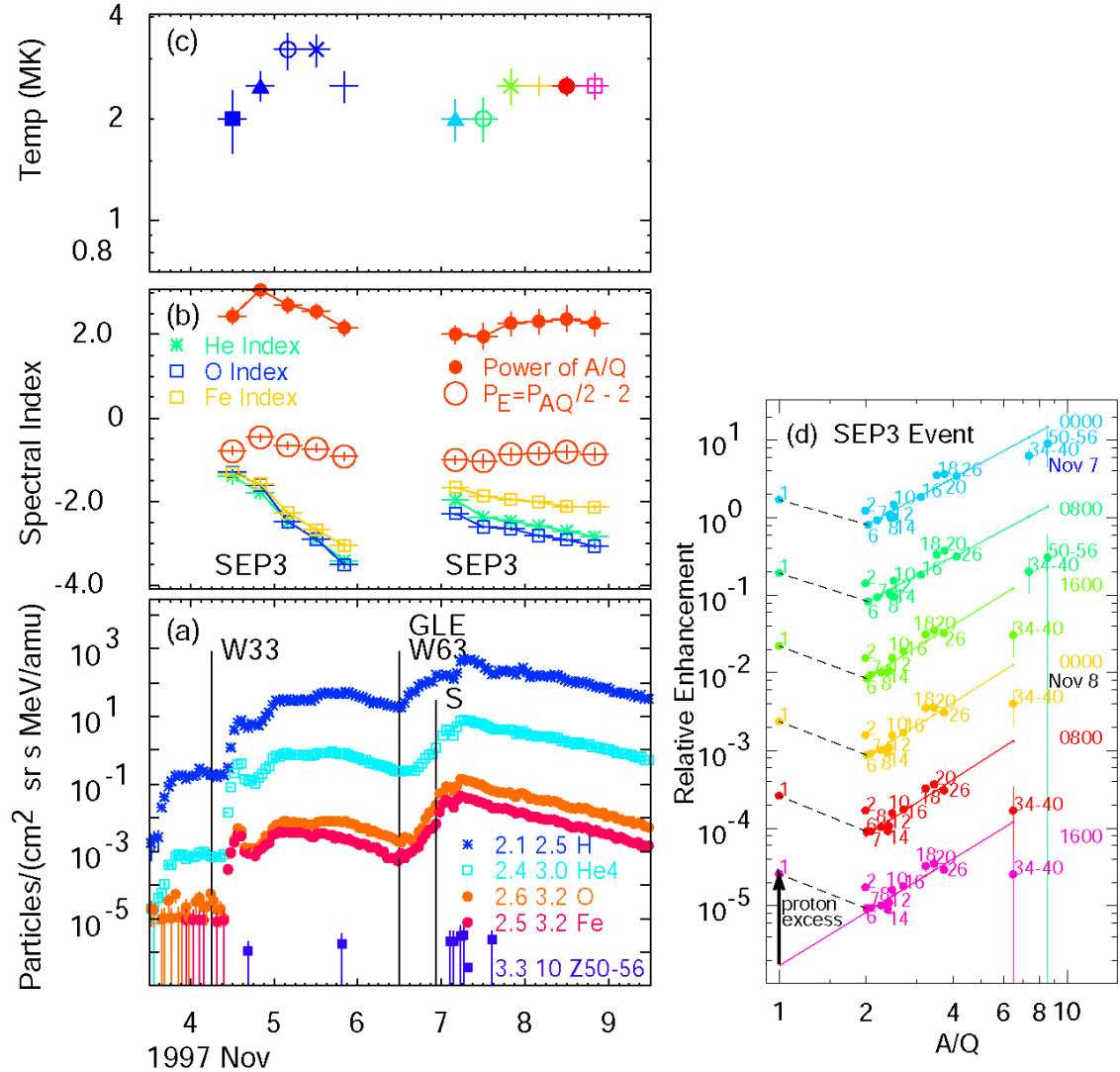

**Figure 5** For two SEP3 events during November 1997 we show (**a**) element time profiles, (**b**) observed spectral indices for He, O, and Fe compared with the power of *A/Q* (solid red circle) and the expected spectral derived from Equation 2 (open red circles), (**c**) derived plasma temperatures, and (**d**) best-fit power-law abundances vs. *A/Q* at each time-interval during the second SEP3 event.

The two events in Figure 5 also share commonality of origin. During the 2 days and 6 hours between the event at 05:58 UT on 4 November at W33 and that at 11:55 UT on 6 November at W63, the Sun rotates just 30º. The SEP remnants of the first events may also be captured to produce seeds for the second. The abundances and powers of *A/Q* in the second event seem to follow on from the first, but the spectra have hardened in the second event and remain so.

Figure 6 shows data for two other related SEP3 events, both GLEs. Again, the SEP3 GLE at W15 on 2 May and the SEP3 GLE at W64 are separated by 3.45 days and 49⁰ while the Sun has rotated ≈ 46⁰, presenting enhanced residue from the first event to the shock in the second. In fact, another significant impulsive event early on 4 May contributes additional extremely Fe-rich material from the source region, now at W34, near mid-course in its rotation. The contribution from this intermediate event is seen after the shock passage on 4 May, especially in the Fe increase, but also in the power of *A/Q* and the spectral indices of He and O.

The expected powers of *E* based upon Equation 2 and shown as large open circles in Figure 6b are far from the observed spectral indices, as is typical for SEP3 events, i.e. the spectra and abundances are unrelated. The two GLEs in Figure 6 have order-of-magnitude proton excesses (not shown); for the SEP2 event at W34, protons are obscured by background, but the event is associated with a 649 km s$^{-1}$ CME. SEP2 and SEP3 events surely contribute some of their excess suprathermal protons to these pools of suprathermal ions so in subsequent events it is impossible to say which of the excess protons came from the suprathermal ions and which came from the ambient plasma.

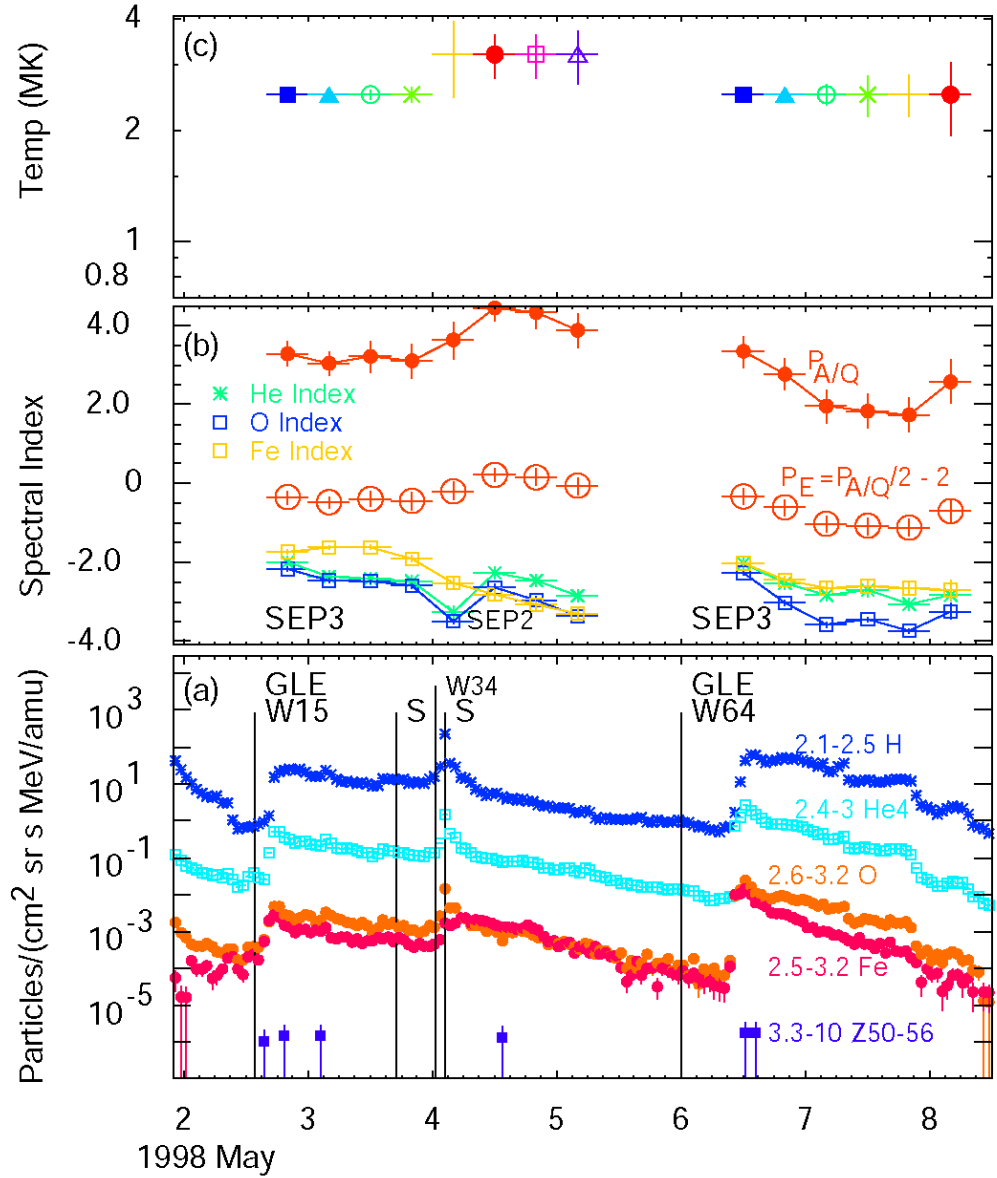

**Figure 6** For two SEP3 events that are GLEs during May 1998 we show (**a**) element time profiles, (**b**) observed spectral indices for He, O, and Fe compared with the power of *A/Q* (solid red circle) and the expected spectral derived from Equation 2 (open red circles), (**c**) derived plasma temperatures,

A final example of SEP3 events in Figure 7 shows another related pair of events that are both GLEs in April 2001. Here the source region has rotated beyond the solar limb by the time of the second event, yet it is a GLE. Both events have ten-fold proton excesses (not shown). However in Figure 7b the expected power of *E* from Equation 2, shown as open circles, seems to agree with the He spectral index for the second event, perhaps coincidentally, since the He index has hardened appreciably for this behind-the-limb event.

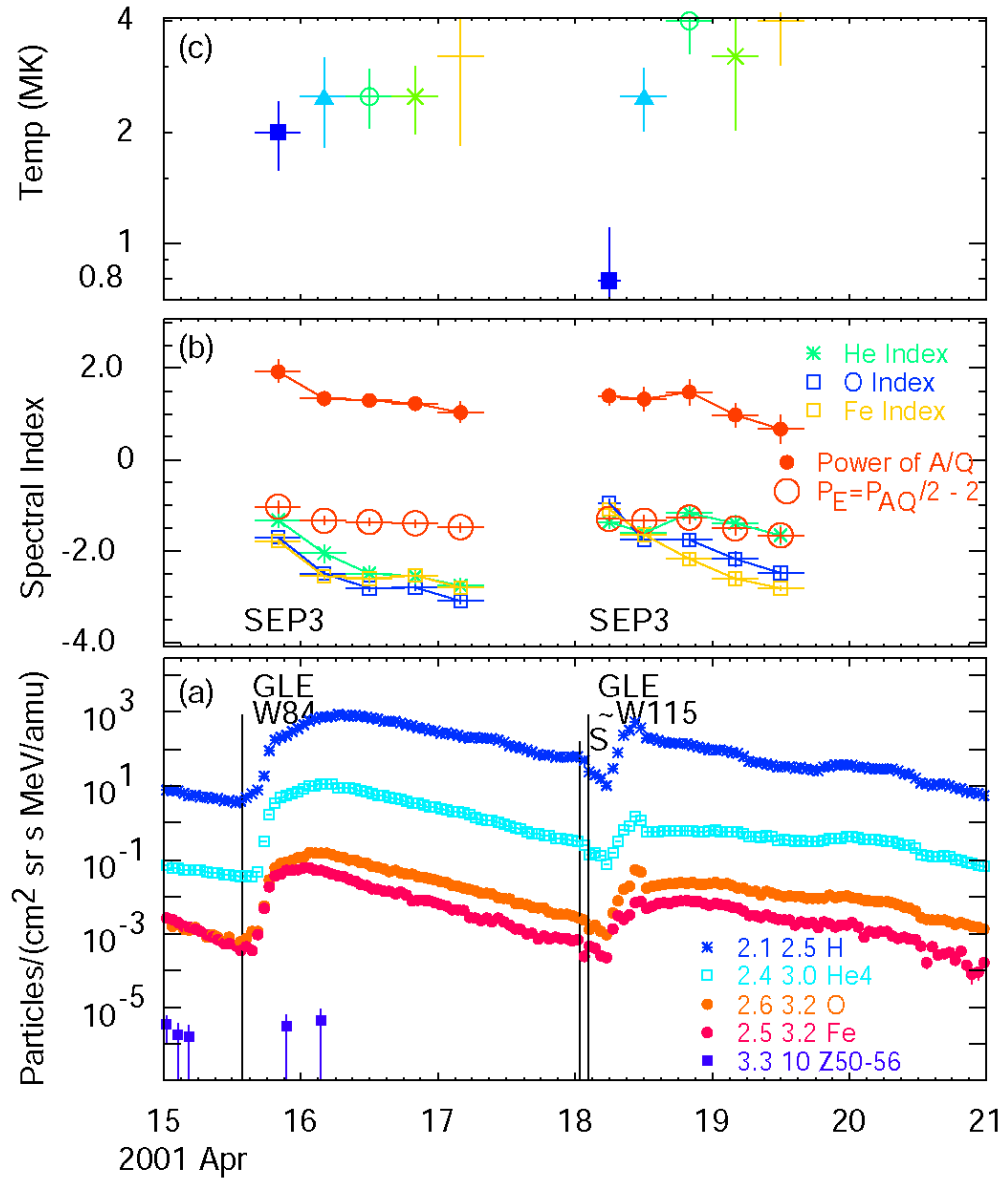

**Figure 7** For two SEP3 events during April 2001, both GLEs, we show (**a**) element time profiles, (**b**) observed spectral indices for He, O, and Fe compared with the power of *A/Q* (solid red circle) and the expected spectral indices derived from Equation 2 (open red circles), and (**c**) derived plasma temperatures.

Clearly, Figures 5, 6, and 7 show that the propensity of SEP3 events to "double-dip" into pools of pre-enhanced impulsive ions is quite common.

## 5. The Pattern of Spectral Indices vs. Powers of *A/Q*

We now compare the spectra and abundance for all classes of SEP events. A previous article (Reames 2021a) compared 8-hour intervals during gradual SEP events and distin-guished SEP3 and SEP4 events by their source plasma temperatures. Here we plot event-averaged spectral indices vs. powers of *A/Q*, so that we can compare events on an equal basis. We also use proton excess >5 (and the standard deviation) to distinguish SEP2 events, which have shock reaccelerated ions, from SEP1 events which do not (only events with measurable proton intensities are included). The factor of 5 allows for significant fluctuations, yet distinguishes the events with order-of-magnitude proton excesses. We will see below that this choice of proton excess seems to agree well with our previous use of CME speed in distinguishing SEP1 and SEP2 events. The resultant distribution of SEP events is shown in Figure 8.

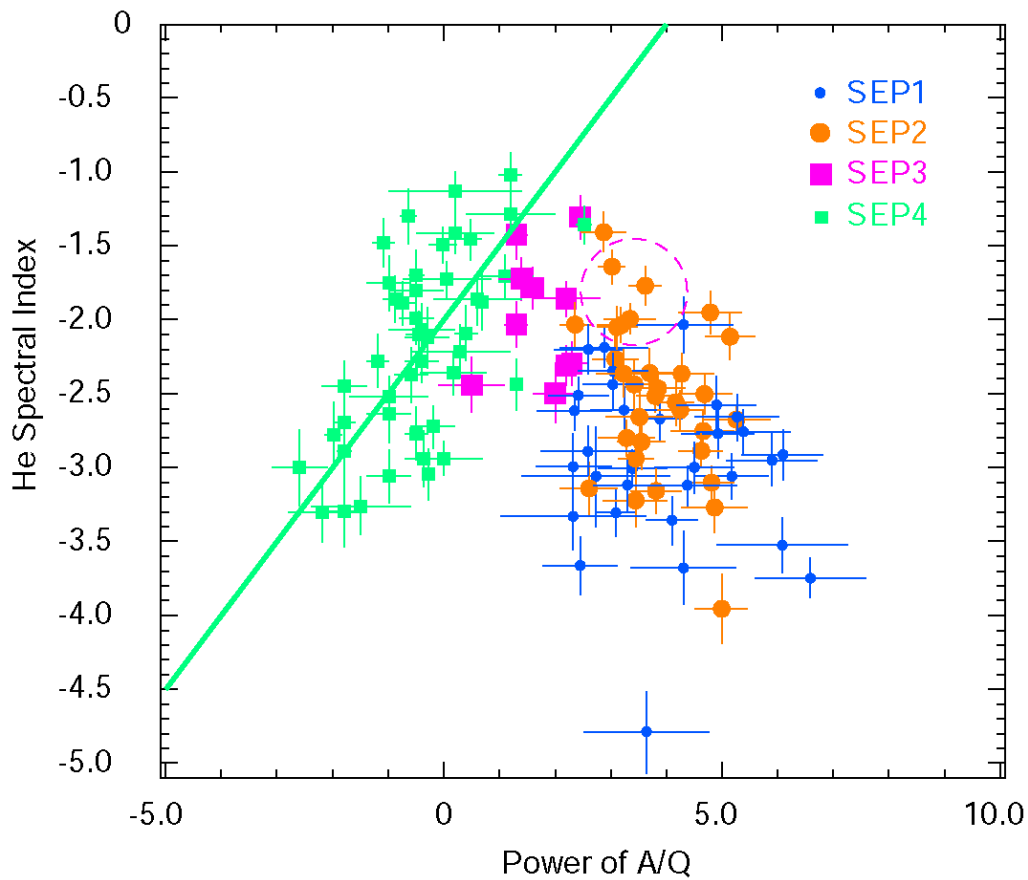

**Figure 8** The distribution of various SEP events is shown in a space of He spectral index vs. abundance-enhancement power of *A/Q*, i.e. *b* vs. *a* from Equation 1   Event classes listed in Section 1 are distinguished by symbols and colors as shown (see text). The green line represents the track defined by Equation 2. The dashed circle surrounds probable additional SEP3 events (see text).

## *5.1 Shock Waves Distinguish SEP2 from SEP1*

Evidence suggests that proton excess does distinguish SEP2 events with dominant proton acceleration from the ambient plasma and heavier ions from the pre-accelerated SEP1 impulsive material, see e.g. Figure 6 of Reames (2021c). However, how much do we allow for underlying variation in the H abundance? Is the choice of the threshold value of proton excess ≈ 5 reasonable? We know that gradual SEP events involve shocks. Is it reasonable that over half of the impulsive events also involve reacceleration of the ions by shock waves? Some justification of this choice is found in Figure 9 which shows a typical average He intensity vs. CME speed and vs. X-ray intensity for those impulsive events that have both measured protons and associated CMEs and/or X-rays.

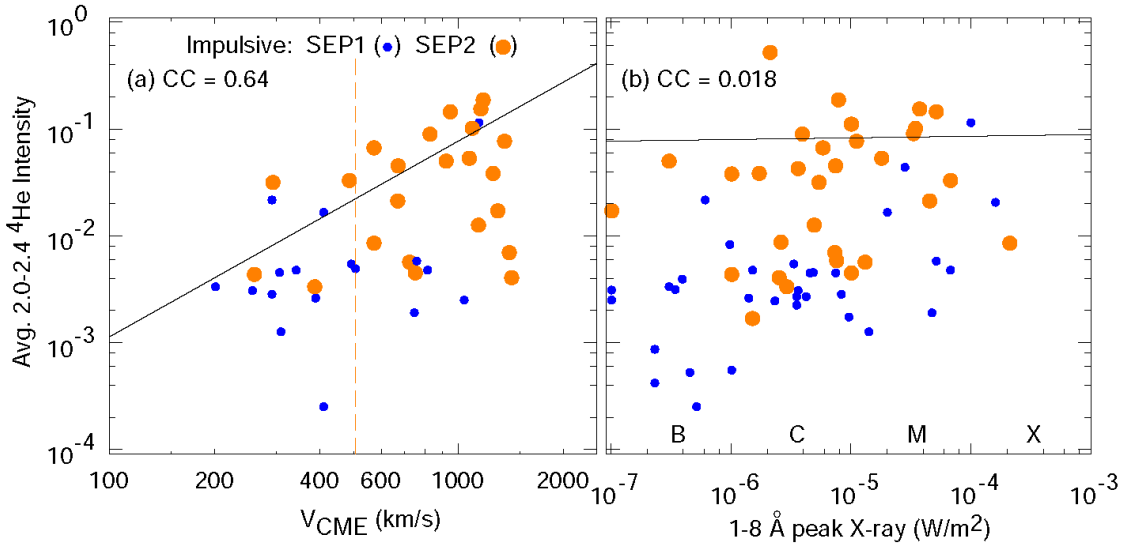

**Figure 9** The distribution of impulsive SEP events in the space of He intensity vs. (**a**) CME speed and (**b**) soft X-ray intensity, with SEP2 events defined by a proton excess factor > 5. Least-squares fit lines are shown. The impulsive SEP events show a 64% correlation (CC stands for cross correlation) with shock speed (especially noticeable for the SEP2 events with proton excesses), not unlike that seen for gradual SEP events, and only a 1.8% correlation with X-ray intensity. The number of events in the two panels differs since fewer events had associations with CMEs than with X-rays bursts, perhaps a sensitivity issue.

First, notice that the overall correlation coefficient for these impulsive SEP events with shock speed in Figure 9a is 64% while that with soft X-ray intensity in Figure 9b is only 1.8%. While that does not prove the importance of shock acceleration, the correlation does compare with similar correlation coefficients with CME speed seen for gradual SEP events (e.g. Kahler, 2001; Kouloumvakos et al., 2019) and it is consistent with our interpretation that a large proportion of impulsive events are SEP2 events that involve shock acceleration.

Second, the dashed line in Figure 9a at $\approx$ 500 km s$^{-1}$ seems to divide the space into SEP1-dominated population below 500 km s$^{-1}$ and a SEP2-dominated population above. Shock speeds of $\approx$ 500 km s$^{-1}$ have long been credited with the nominal beginning of shock acceleration (e.g. Reames, Kahler, and Ng, 1997). Occasionally, SEPs are measured with associated CME speeds as low as 300 km s$^{-1}$ (Kahler, 2001). This also conforms to the onset of the shock speed distribution of in situ interplanetary shocks (Reames 2012). By itself, the correlation of intensity with CME speed might only mean that larger, more energetic jets have faster CMEs, but the presence of a second component, excess protons, beginning at an appropriate speed for onset of shock acceleration, supports our division of SEP1 and SEP2 events as the onset of shock acceleration. The onset of a second component, signaled by a significant proton excess, is consistent with the onset of a second acceleration mechanism, e.g. a shock.

The important identification of proton excesses and SEP2 and SEP3 events depends upon comparing H to the trend of the ions with $Z > 2$, rather than comparing H/He, for example. The abundance of He, or He/C or He/O, seems to be partly affected by FIP-dependent factors arising from the uniquely high FIP = 24.6 eV of He (e.g. Reames, 2017, 2021d; Laming. 2009, 2015). While H/He is an obvious comparison of dominant elements, studying H/He alone mixes the factors that modify H with those that modify He making them very difficult to disentangle.

## 6. A Summary

It seems appropriate to summarize the various classes of SEP events and the information we use to distinguish them.

(i) Impulsive and gradual SEP events are distinguished by a bimodal abundance pattern seen plotting Ne/O vs. Fe/O (e.g. Figure 1 in Reames, Cliver, and Kahler, 2014a; see also Reames 1988; Reames and Ng 2004). Impulsive events are Fe rich. We usually do not use $^3$He/$^4$He because it is very erratic and varies strongly with energy.

SEP Spectra and Abundances                                          D. V. Reames

(ii)  SEP1 and SEP2 events are distinguished using a proton excess factor > 5 as evidence of shock reacceleration of local SEP1 ions and ambient protons as discussed above (see Figure 6 in Reames, 2021c).

(iii) SEP2 and SEP3 events can be distinguished statistically since SEP2 shocks sample local suprathermals (e.g. $^4$He/C) with large variations, while SEP3 shocks sample pools with averaged abundances (see Figure 8 in Reames, 2021c).

(iv) SEP3 and SEP4 events are distinguished using proton excess factor > 5 and fit temperatures ≥ 2 MK for SEP3 events as evidence of shock reacceleration of pools of pre-enhanced impulsive suprathermal ions plus ambient protons as discussed above.

However, these categories are less to label individual events and more to identify combinations of physical processes and their typical properties.

Consider, now, the events in the dashed red circle in Figure 8. The boundary between SEP2 and SEP3 events that exists in Figure 8 is determined primarily by the criterion (i) above, i.e. that impulsive SEP events are Fe rich, but particles accelerated from pools of accumulated impulsive suprathermal ions are also likely to be Fe rich. Perhaps these SEP3 events have been mistakenly assigned as individual impulsive SEP2 events. A few of the SEP2 events with flat He spectra (index > -2.5) and low powers of *A/Q* (< 4), surrounded by the dashed red circle in Figure 8, also have the characteristically confined C/$^4$He ratio typical of the SEP3 events (see e.g. Figure 8 in Reames, 2021c) and may be incorrectly identified as SEP2 events. The power of *A/Q* for the impulsive heavy-ion enhancements need not be diminished in SEP3 events, but dilution by contribution to the suprathermal pools from gradual SEP events is also possible.

From the individual events we have seen that the prediction of Equation 2 seems to follow the complex spectral and abundance variations in the shock-dominated SEP4 events in Figures 2, 3, and 4, while successive CME-driven shocks often sample the same pool of jet-fed impulsive suprathermal ions that can persist for several days as active regions rotate around the Sun, as seen in Figures 5, 6, and 7.

Shock acceleration is common and important, even in impulsive SEP events, where over half of the Fe-rich events we see are SEP2 events with intensities of the SEP1 ions boosted by shocks and added protons derived from the ambient plasma.

According to shock theory, the energy spectral index of the accelerated ions is related to the shock compression ratio (e.g. Jones and Ellison, 1991). When Equation 2 is valid, this means that the ion abundance enhancements are also determined by the shock compression ratio, i.e. in SEP4 events the injection of heavy ions into the shock from the coronal plasma may depend upon *A/Q* of the ion to a power that depends only upon the shock compression ratio.

In Figure 8, we can see that impulsive SEP events are widely distributed about nominal values of $a \approx +4$ and $b \approx -3$ in Equation 1. Theoretically, Drake et al. (2009) found that magnetic reconnection supports the observed average power law in *A/Q* near +3, but do not discuss the spectral index. Recently Che, Zank, and Benz (2021) found evidence during magnetic reconnection for a decreasing broken power-law spectral index that is quite steep (e.g. $b \approx -7.5$) above ~1 MeV amu$^{-1}$, but they do not discuss element abundances. It would be interesting to understand any physical relationship between spectra and abundances and their large variation in these events.

## Data Availability

Wind/LEMT data on ion spectra and abundances are available at https://cdaweb.gsfc.nasa.gov/sp_phys/.

## Disclosure of Potential Conflicts of Interest

The author declares he has no conflicts of interest.